\documentclass[%
aip,
amsmath,amssymb,
reprint,%
]{revtex4-1}
\DeclareUnicodeCharacter{2212}{-}
\usepackage{graphicx}
\usepackage{dcolumn}
\usepackage{bm}
\usepackage{comment} 

\usepackage[utf8]{inputenc}
\usepackage[T1]{fontenc}
\usepackage{mathptmx}
\usepackage{etoolbox}

\usepackage[english]{babel}

\usepackage{siunitx}
\usepackage{import}
\usepackage[hidelinks]{hyperref}
\usepackage[capitalise]{cleveref} 
\usepackage{xcolor}

\makeatletter
\def\@email#1#2{%
	\endgroup
	\patchcmd{\titleblock@produce}
	{\frontmatter@RRAPformat}
	{\frontmatter@RRAPformat{\produce@RRAP{*#1\href{mailto:#2}{#2}}}\frontmatter@RRAPformat}
	{}{}
}%
\makeatother

\begin{document}
	\title{Simple, highly-stable transfer cavity for laser stabilization based on a carbon-fiber reinforced polymer spacer}
	
	\author{T. Zwettler}
	\affiliation{Institute of Physics, \'Ecole Polytechnique F\'ed\'erale de Lausanne (EPFL), CH-1015 Lausanne, Switzerland}
	\affiliation{Center for Quantum Science and Engineering, Ecole Polytechnique F\'ed\'erale de Lausanne, CH-1015 Lausanne, Switzerland}
	
	\author{Z. Xue}
	\affiliation{Institute of Physics, \'Ecole Polytechnique F\'ed\'erale de Lausanne (EPFL), CH-1015 Lausanne, Switzerland}
	\affiliation{Center for Quantum Science and Engineering, Ecole Polytechnique F\'ed\'erale de Lausanne, CH-1015 Lausanne, Switzerland}
		
	\author{G. Bolognini}
	\affiliation{Institute of Physics, \'Ecole Polytechnique F\'ed\'erale de Lausanne (EPFL), CH-1015 Lausanne, Switzerland}
	\affiliation{Center for Quantum Science and Engineering, Ecole Polytechnique F\'ed\'erale de Lausanne, CH-1015 Lausanne, Switzerland}
	
		\author{T. Bühler}
	\affiliation{Institute of Physics, \'Ecole Polytechnique F\'ed\'erale de Lausanne (EPFL), CH-1015 Lausanne, Switzerland}
	\affiliation{Center for Quantum Science and Engineering, Ecole Polytechnique F\'ed\'erale de Lausanne, CH-1015 Lausanne, Switzerland}
	
	\author{L. Hruby}
	\affiliation{Institute for Quantum Electronics, Eidgenössische Technische Hochschule Zürich, CH-8093 Zürich, Switzerland}

    \author{A. Fabre}
	\affiliation{Institute of Physics, \'Ecole Polytechnique F\'ed\'erale de Lausanne (EPFL), CH-1015 Lausanne, Switzerland}
	\affiliation{Center for Quantum Science and Engineering, Ecole Polytechnique F\'ed\'erale de Lausanne, CH-1015 Lausanne, Switzerland}

 	\author{T. Donner${}^*$}
	\affiliation{Institute for Quantum Electronics, Eidgenössische Technische Hochschule Zürich, CH-8093 Zürich, Switzerland}
    \email[]{donner@phys.ethz.ch}
    
	\author{J.-P. Brantut${}^{**}$}
	\affiliation{Institute of Physics, \'Ecole Polytechnique F\'ed\'erale de Lausanne (EPFL), CH-1015 Lausanne, Switzerland}
	\affiliation{Center for Quantum Science and Engineering, Ecole Polytechnique F\'ed\'erale de Lausanne, CH-1015 Lausanne, Switzerland}
    \email[*]{jean-philippe.brantut@epfl.ch }

	\begin{abstract}
        We describe the design and operation of a high-stability Fabry-Perot cavity, for laser stabilization in cavity quantum-electrodynamics experiments. Our design is based on an inexpensive and readily available uniaxial carbon-fiber reinforced polymer tube spacer, featuring an ultra-low thermal expansion coefficient. As a result, our $\SI{136}{\milli\meter}$-long cavity, which has a finesse of ${5160}$, shows a coefficient of thermal expansion of $1.6 \times 10^{-6}~\mathrm{K}^{-1}$. Enclosing it in a hermetic chamber at room-pressure and using a simple temperature stabilization, we observe absolute frequency excursions over a full day below $50~\mathrm{MHz}$ for a laser operating at $\SI{446.785}{\tera\hertz}$. The frequency stability is limited by the imperfect thermal isolation from the environment and can be corrected using a built-in piezo-electric actuator. In addition, we discuss a different variant of this design and identify future improvements. Our system provides a cost-effective and robust solution for transferring laser stability over different wavelengths, as well as for linewidth reduction or spectral filtering of CW laser sources for applications in quantum science. 
	\end{abstract}
	
\maketitle

\section{Introduction}
Fabry-Pérot etalons and reference cavities are widely used in laser stabilization. High-end, ultra-low expansion reference cavities are the workhorse of atomic clocks \cite{NicholsonSEOAAC2015, BothwellRTGRAAMSAS2022} and are also employed in quantum metrology \cite{PedrozoEOAOACT2020} as well as gravitational wave interferometers \cite{KweeSHPLSFTGWDALIGO2012}, where Hz-scale stability of lasers can be reached. In contrast, robust and inexpensive monolithic cavities with a spectral resolution of GHz are used for filtering applications in quantum optics \cite{PalittapongarnpimMFCFEIQO2012}. In the intermediate regime of the tens of kHz to MHz resolution, atomic spectroscopy allows for absolute frequency stabilization of lasers  \cite{PearmanPSOACATATLFL2002, BjorklundFMS1983, CorwinFSDLWTZSIAV1998}, but transferring this performance throughout the entire optical spectrum requires a transfer cavity with performance in a similar range of linewidths \cite{ThorpeLFSACTOSLTOC2008, SchuenemannSSFTFOLOTL1999}. Furthermore, this type of cavity is particularly useful for spectral filtering \cite{GalinskiyPCTOAUMRNIMGS20, DideriksenRTSPSWNMBIM2021} and narrowing of lasers \cite{AlnisSLDLBTVATCUEGFP2008, SchoofRTLOADL2001} for applications in quantum science, atomic and molecular physics experiments. However, achieving long-term stability and high-degree of noise rejection, while not requiring the expensive technologies used for ultra-low expansion cavities operated in atomic clocks, implies careful engineering and can represent a substantial technical and cost barrier.
	 
In this paper, we present the design and operation of a simple and flexible Fabry-Pérot (FP) cavity, which is used as the frequency discriminator in a transfer cavity locking technique, ensuring the relative stability of two lasers at vastly different wavelength. In this stabilization scheme, a mode of a FP cavity is locked to a master laser and the slave laser frequency is then in turn stabilized to another mode of the cavity. The passive frequency stability of the FP cavity ensures the relative stability of the two lasers as well as naturally provides the possibility of linewidth reduction of the slave laser via Pound-Drever-Hall locking \cite{AlnisSLDLBTVATCUEGFP2008, SchoofRTLOADL2001}. The core piece is an inexpensive spacer made of a standard uniaxial carbon-fiber reinforced polymer (CFRP) tube. Uniaxial CFRP features an extraordinarily low coefficient of thermal expansion (CTE) of $\SIrange{1}{2}{\times 10^{-6}\kelvin^{-1}}$ along the fiber direction \cite{DongEOTECOCFRCUESI2018, PradereTALCOTEOCFAHT2008}, similar to standard-grade invar or fused silica \cite{CallisterMSAE2020}. Our groups have developed several generations of this system deployed in our cavity quantum-electrodynamics (cQED) experiments \cite{HrubyMasterThesis2013, LandigPhDThesis2016, RouxPhDThesis2022}, showing long-term robustness and easy operation. Here we present the latest generation, which is operated at EPFL as a transfer cavity between \SI{1064}{\nano\meter} and \SI{671}{\nano\meter}. It features a CTE of $1.6 \times 10^{-6}~\mathrm{K}^{-1}$ for a cavity length of $\SI{136}{\milli\meter}$, allowing for thermal tuning of the cavity at a rate of \SI{700}{\mega\hertz/\kelvin}. This is ideal for coarse tuning, while a simple thermal stabilization technique yields a hours-scale stability of the order of \SI{50}{\mega\hertz} for a laser operating at $\SI{446.785}{\tera\hertz}$. The paper is structured as follows: in the first part, we describe in detail the mechanical design of the cavity as well as its enclosure. In the second part, we present the performance of the system. We conclude by presenting a variant of the design and extensions that we have either tested or considered, allowing for tailoring the design to a specific application.
	
	\begin{figure}[h]
		\includegraphics[width=\linewidth]{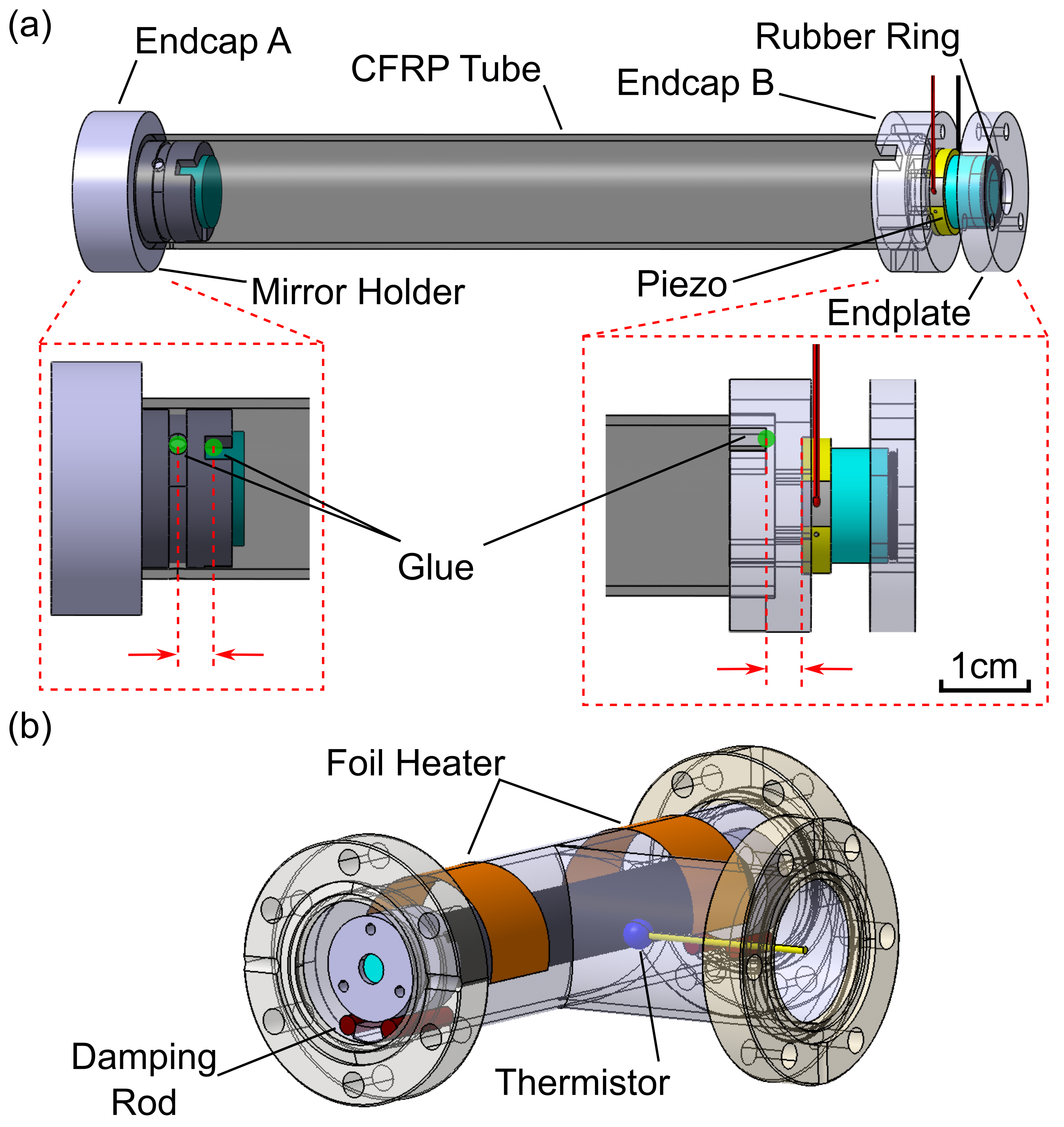}
		\centering
		\caption{\textbf{Transfer cavity design.} (a) Schematic of the FP cavity, consisting of a CFRP spacer and two stainless steel endcaps (A, B) holding the mirrors (details in insets). The endcaps are designed to cancel the stainless steel contribution to thermal expansion of the cavity. The cavity spacer and endcaps are connected with an epoxy glue (light green dots), which is only applied orthogonally to the cavity axis. Whereas the mirror in endcap A is glued (light green dot), the endcap B contains a ring piezo (yellow) for fine-tuning the cavity length. The mirror of endcap B is pressed against the piezo using a stainless steel endplate and a rubber O-ring with screws (not shown). (b) The cavity is placed in a sealed vacuum T-piece for environmental isolation. The vacuum T-piece is temperature regulated using two foil heaters as actuators and a thermistor placed inside the sealed chamber.}
		\label{fig1}
	\end{figure}

\section{Design}

Our cQED experiments, where atoms are coupled to a high-finesse optical resonator, typically involve cavities with linewidths of the order of \SIrange{0.1}{1}{\mega\hertz} \cite{MivehvarCQEDWQG2021}, often limited by losses of commercially available high-reflectivity mirrors at the relevant wavelengths \cite{RempeMOULLIAOI92, RudelisDODCMIUHV2023}. To stabilize laser frequencies on short time-scales to the same order of magnitude as the linewidth of a high-finesse cavity, while providing relative stability for lasers ranging from near-resonant to far-detuned from atomic resonance, doubly-resonant transfer cavities with comparable linewidths are needed. To achieve such narrow linewidths without relying on very high finesse, which would require costly and scarcely available super-mirrors, we utilize transfer cavities with a large mirror spacing. As a result, the primary challenge is the long-term drift of the cavity length, which can potentially exceed the compensation range of the piezo-actuators, eventually causing a loss of lock in the entire laser system. These drifts primarily arise from fluctuations in ambient temperature and humidity. While humidity-induced drifts can be effectively mitigated by housing the resonator in an enclosed volume, minimizing thermal drifts presents a greater challenge. To address this, we employ a CFRP tube as spacer between the cavity mirrors and place the entire cavity in a temperature-stabilized enclosure.

More precisely, we focus on a cavity using identical plano-concave mirrors with a reflectivity specified at $\SI{99.95(2)}{\%}$ at both $\SI{671}{nm}$ and $\SI{1064}{\nano\meter}$ and radius of curvature of $\SI{1.00(1)}{\meter}$ (Optoman). The specified reflectivity enables the Finesse to reach $\num{6(3)} \times 10^3$. Together with the chosen cavity length of $\SI{135(1)}{\milli\meter}$, this results in a  free spectral range (FSR) of $\SI{1.11(1)}{\giga\hertz}$ and a target cavity full-width-half-maximum (FWHM) of $\SI{180(70)}{\kilo\hertz}$. The cavity alignment is facilitated by being far from either concentric or confocal limit \cite{HauckMSOOR1980}. 

Our design is presented in Fig.~\ref{fig1}. The spacer is a readily available tube of pultruded, unidirectional carbon-fiber-reinforced polymer with a length of $\SI{140.15}{\milli\meter}$. Endcaps holding the mirrors and piezo actuator are made of stainless steel for easy machinability. Given that stainless steel has a CTE that is an order of magnitude greater than that of carbon fibers \cite{CallisterMSAE2020}, we compensate for its impact on the cavity's thermal expansion by mounting the mirror on endcap A on the inner side and the mirror on endcap B on the outer side of the stainless steel section, which has thickness of $\SI{4}{\milli\meter}$ and contributes to the cavity spacer (see insets Fig.~\ref{fig1}). This configuration results in a common-mode motion due to thermal expansion of both mirrors without affecting the overall cavity length. The endcaps and the spacer are glued together using an epoxy glue (Araldite RAPID epoxy glue).  Because the glue has a very high CTE of  $(5\sim9)\times10^{-5}\ \mathrm{K}^{-1}$ \cite{Silva_HandbookofAdhesionTechnology2018}, we only apply it in the direction orthogonal to the cavity axis, at the points indicated in light green in Fig.~\ref{fig1}. Endcap B also contains a piezoelectric ring chip (Thorlabs PA44M3KW), which has a maximum displacement of $\SI{3.9}{\mu\meter}$ at an input voltage of $\SI{150}{V}$. This allows to stabilize the length of the cavity onto a reference laser. The piezo is electrically isolated from the inner part of endcap B by a circular groove. As we show below, this maximum displacement is sufficient to compensate residual thermal drifts over the long term. 

For applications as a transfer cavity for different wavelengths, variations of the density of air induce relative changes of the refractive index, yielding differential frequency drifts in the stabilization system \cite{RiedleSAPCOACWDFSBUASTC1994}. We house the cavity in a T-shaped, hermetically sealed stainless steel ConFlat chamber, equipped with standard, antireflection-coated viewports and electrical feedthroughs. This enforces constant density of air inside the cavity, yielding an insensitivity to ambient humidity, pressure or temperature changes to lowest order, without the extra cost and burden of a permanently attached vacuum pumping unit. The chamber is also providing a significant thermal and acoustic isolation barrier. To isolate from vibrations originating from the optical table, such as moving shutters, the cavity rests on two pairs of viton rods. For temperature stabilization, a thermistor is loosely attached to the body of the cavity spacer, while flexible foil heaters (Thorlabs HT10K) are mounted on the exterior of the chamber to serve as actuators.

The cavity is assembled using kinematic V-clamp mounts (Thorlabs KM200V/M) to hold endcaps A and B in place during alignment and glueing. Initially, a coarse alignment of the cavity is performed without the CFRP spacer. Next, the spacer is inserted by mounting one of the kinematic V-clamp mounts on a single-axis translation stage. Fine-alignment is then carried out, and the endcaps are glued to the spacer. Finally, the clamps are removed once the adhesive has set.
    
\section{Performance}

	\begin{figure}[h]
		\includegraphics[width=\linewidth]{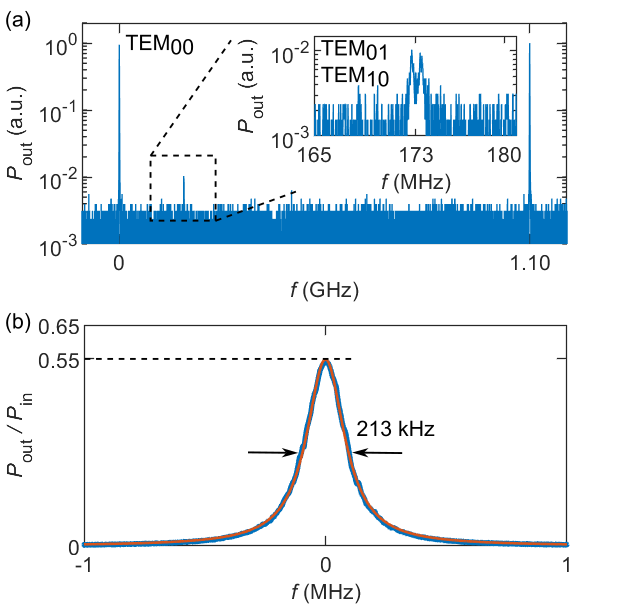}
		\centering
		\caption{\textbf{Transmission spectrum.} (a) Transmission spectrum (blue) at $\SI{1064}{nm}$ over the range of one FSR. The $\text{TEM}_{01}$ and $\text{TEM}_{10}$ modes shown in the inset are almost degenerate, indicating good rotationally symmetric cavity alignment. (b) Transmitted power $P_{\mathrm{out}}$ of the $\text{TEM}_{00}$ mode (blue) normalized to the input power $P_{\mathrm{in}}$ with a Lorentzian fit (orange).}
		\label{fig3}
	\end{figure}

We characterize the transmission spectrum of the cavity using a narrow-linewidth $\SI{1064}{\nano\meter}$ laser (NKT Koheras Adjustik). First, we scan the laser frequency by about one FSR while keeping the cavity length fixed, as can be seen from Fig.~\ref{fig3} (a). The frequency axis $f$ is calibrated by applying an amplitude-modulation to the laser at a known frequency using a fibered electro-optic amplitude modulator and using the known sideband frequency as an absolute reference. We obtain a FSR of $\SI{1.10}{\giga\hertz}$ corresponding to a cavity length of $L_0 = \SI{136}{\milli\meter}$. Then, we perform a narrower frequency scan around a single cavity resonance by keeping the laser frequency fixed and changing the cavity length using the piezoelectric actuator. We extract the full-width-half-maximum (FWHM) of the cavity using a Lorentzian fit, as can be seen from Fig.~\ref{fig3} (b). From the fit, we obtain a FWHM of $\SI{213(3)}{\kilo\hertz}$ yielding a finesse of $\SI{5.16(8)}{\times10^{3}}$. 

We achieve a very good spatial coupling to the $\mathrm{TEM}_{00}$ mode with a suppression of the first order transversal $\mathrm{TEM}_{01, 10}$ modes by a factor $100$. The ratio of cavity input to maximum output power is $54.9\%$. Using the measurement of the transmitted power and the finesse, we estimate the transmissivity to be $\SI{451(7)}{ppm}$, which agrees well with the specified reflectivity given by the manufacturer and is presumably limited by mirror losses \cite{HrubyMasterThesis2013, RempeMOULLIAOI92}. The very small frequency difference between the two first-order transverse modes indicates a small astigmatism that can be interpreted as a \SI{0.8}{\%} relative difference between the mirror radii of curvatures along two directions, in good agreement with the mirror specifications. We did not observe any polarization dependence in the cavity transmission.

We now characterize the thermal properties of the assembly by measuring its effective CTE.
Given a temperature change $\Delta T$, a spacer material with a linear effective CTE $\alpha$ will have a relative change of its length by $\Delta L/L_0$ and a relative change of its cavity resonance frequency $\Delta \nu/\nu_0$ according to \cite{RiedleSAPCOACWDFSBUASTC1994}:

\begin{equation}
    \frac{\Delta\nu}{\nu_0} \approx -\frac{\Delta L}{L_0} = - \alpha \Delta T.
    \label{eqn1}
\end{equation}

	\begin{figure}[h]
		\includegraphics[width=\linewidth]{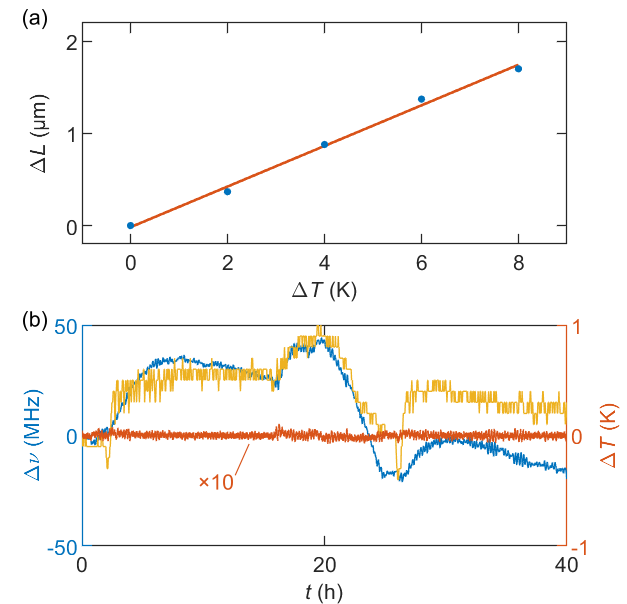}
		\centering
		\caption{\textbf{Thermal expansion and long-term stability.}(a) Change of cavity length $\Delta L$ as a function of a change of temperature $\Delta T$ (blue circles) from which the CTE is obtained using a linear fit (solid orange line). (b) Drift of the cavity resonance frequency $\Delta \nu$ over forty hours, while the temperature of the enclosure is actively stabilized. The right axis shows temperature changes $\Delta T$ during this period of both the room (light orange) and the sensor used for temperature stabilization in the middle of the chamber (dark orange). While the sensor indicates a stable temperature, the frequency drifts are clearly correlated with changes of the room temperature, suggesting that they are caused by temperature gradients.}
		\label{fig2}
	\end{figure}
	
To obtain the cavity resonance frequency, we stabilize the laser to the cavity resonance using a side-of-fringe locking technique and measure the laser frequency using a wavemeter (HighFinesse WS6) with a frequency resolution of a few MHz. As can be seen from Fig. \ref{fig2}(a), the change in cavity length $\Delta L$ (blue circles) is linear with temperature, from which we extract the cavity length expansion rate of $\SI{0.22}{\mu \meter/\kelvin}$ using a linear fit (orange). This results in a CTE of $\SI{1.6}{\times10^{-6}\kelvin^{-1}}$, which is on the same order as the bare CFRP tube, indicating that the contribution of the stainless steel endcaps to the thermal expansion of the cavity body is well cancelled.

The long-term stability of the cavity relies on its temperature stabilization and the isolation from the environment through the sealed chamber. We set the cavity temperature to be $\SIrange{1}{2}{\kelvin}$ above the room temperature and monitor the resonance frequency of the cavity using a laser at $\SI{671}{nm}$ and a another wavemeter (HighFinesse WS8), providing higher frequency resolution (see Fig.~\ref{fig2} (b)). The temperature in the middle of the sealed chamber is monitored using a thermistor and stabilized to below $\SI{\pm10}{\milli\kelvin}$ (dark orange). However, we still observe a frequency drift of up to $\SI{50}{\mega\hertz}$, which is clearly correlated with a change of the room temperature outside the chamber (light orange). We attribute this sensitivity of the cavity to the environment to temperature gradients inside the chamber. However, the temperature stabilization suppresses more than $90\%$ of the cavity thermal expansion caused by the room temperature drift, compared to the frequency shift rate of the cavity resonance of $\SI{700}{\mega\hertz/\kelvin}$ at $\SI{446.785}{\tera\hertz}$. 

\section{Alternative Design}

\begin{figure}[h]
		\includegraphics[width=\linewidth]{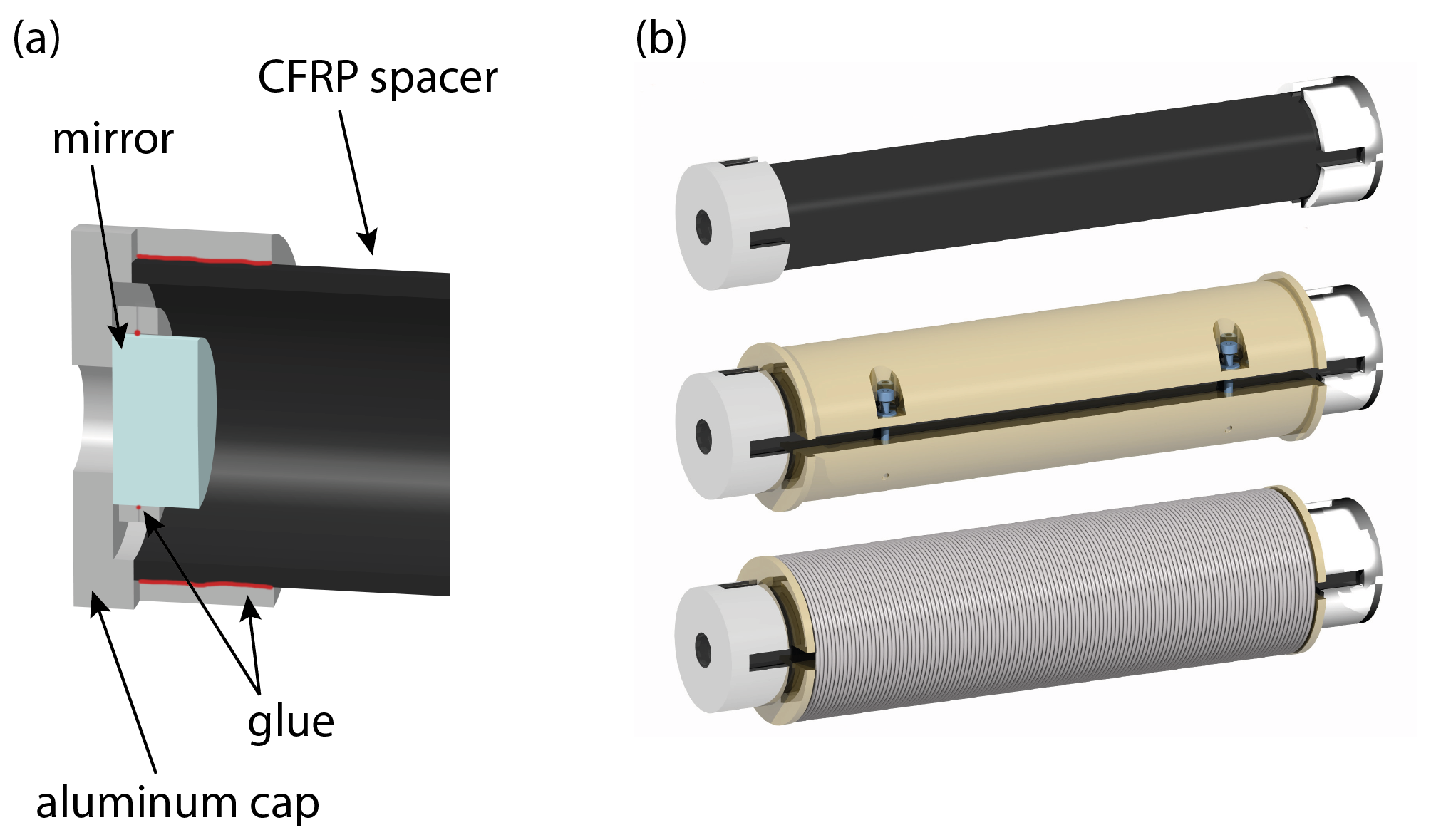}
		\centering
		\caption{\textbf{Alternative design.} (a) Detail of the aluminum endcap holding the mirror, where the thermal expansion is compensated by positioning the glue point (red) in the plane of the CFRP spacer's end. (b) The assembled cavity (top) is in close thermal contact with brass shells (middle), around which a heating wire is wound (bottom).}
		\label{fig4}
	\end{figure}

Based on the same design principle using a CFRP cavity spacer, we developed an alternative configuration~\cite{HrubyMasterThesis2013} currently in operation at ETH, as shown in Fig.~\ref{fig4}. This system serves as a transfer cavity while simultaneously being used to narrow the linewidth of a diode lasers locked to it. To increase passive short-term stability, we remove the piezo and rely exclusively on thermal tuning to adjust the cavity resonance, since potential short-term fluctuations in the cavity resonance can be caused by voltage fluctuations driving the piezo actuator.

The mechanical design again passively compensates for the thermal expansion of the endcaps that attach the mirrors to the CFRP spacer. In this configuration, the mirrors are glued to aluminum endcaps at points that they lie in the plane where the CFRP tube terminates, as illustrated in Fig.~\ref{fig4} (a). This cavity has a length of \SI{144.7}{\milli\meter}, and a measured finesse of $2.3\times10^3$, consistent with the mirror specifications. The CFRP spacer is placed in between two brass shells, see Fig.~\ref{fig4} (b). Good thermal contact between spacer and brass shells is ensured by applying thermal paste. A heating wire is wound around these shells in counter-propagating directions such as to minimize induced magnetic fields which could impact nearby components or experiments. The brass shells act as a thermal low-pass filter, reducing the impact of environmental thermal fluctuations and helps creating a uniform temperature distribution across the CFRP spacer (we measure less than $\SI{100}{\milli\kelvin}$ difference between the center and the edge of the brass shell). Together with a temperature sensor placed at the center of the spacer and a custom-built temperature controller, this setup allows precise regulation of the device's temperature. We determine a CTE of $1.7 \times 10^{-6}~\mathrm{K}^{-1}$, which is comparable to the previously described design. This residual thermal expansion enables convenient tuning of the cavity resonance. By providing feedback to the temperature control, we lock the cavity to the analog output of a wave meter (HighFinesse WS 7), achieving stabilization of the cavity resonance for over 12 hours at a set frequency, with a measured in-loop standard deviation of $\SI{1.8}{\mega\hertz}$.

In addition to serving as a thermal low-pass filter, the mass of the brass shells also is part of the vibration isolation: the cavity is resting on Sorbothane shock absorbers, forming a mass-spring system that suits as low pass filter for vibrational excitations from the optical table. The entire assembly further is supported by a $\SI{40}{\milli\meter}$ diameter Teflon post, which is a poor acoustic conductor. These measures were crucial in minimizing the impact of  vibrations. Using high bandwidth lock electronics (Toptica FALC), we achieved a linewidth below $\SI{40}{\kilo\hertz}$ for an external cavity diode laser (Toptica DLpro) locked to this cavity, with an integration time of $\SI{1}{\second}$.

\section{Discussion}

Our work demonstrates how an inexpensive CFRP spacer can be used to realize transfer cavity systems with low thermal drifts. While the performance is far from reaching the stability of high-end reference cavities used for atomic clocks and precision metrology, our design is well suited to match the requirements of typical cavity QED experiments.

There are several ways to improve both the short-term and long-term frequency stability of our design without the addition of expensive elements. A higher robustness against thermal fluctuations of the environment might be reached for the first design discussed by applying a more uniform heating of the sealed ConFlat vacuum chamber to reduce temperature gradients across the cavity body. Furthermore, we observe that an evacuation of the volume containing the cavity is showing an improvement in the frequency stability (on time scales up to $\SI{100}{\second}$), most likely due to a further decoupling of the cavity from the thermal environment. Improvements regarding the short-term stability might be reached by improving the vibration isolation, e.g. by supporting the cavity specifically at its Airy points along the horizontal cavity axis, such that vertical accelerations due to vibrations produce no angular deflections of the mirror faces \cite{PhelpsAPOAMB1966, GalinskiyPCTOAUMRNIMGS20}.

\section{Acknowledgements}
	
We acknowledges funding from the Swiss State Secretariat for Education, Research and Innovation (Grants No. MB22.00063, MB22.00090, and UeM019-5.1), and from the Swiss National Science Foundation SNSF (Grant No. 217124 and 221538). AF acknowledges funding from the EPFL Center for Quantum Science and Engineering.
	
\section{Data availability}
The CAD design used in this work is available via Zenodo \cite{LQGCADTC2024}, while the most up-to-date version can be accessed via our Github repository \cite{LQGCADTCGithub2024}. The experimental data can be accessed at \cite{LQGData2024}.

\section{Author declarations}
The authors have no conflicts to disclose.

\bibliographystyle{naturemag}

\bibliography{transfercavitybibliography}
	
	
\end{document}